\pdfoutput=1
\documentclass[bibyear,traditabstract,rnote]{aa}
\usepackage{graphicx}
\usepackage{txfonts}
\begin{document}
\title{Principal component analysis-based inversion of effective
  temperatures for late-type stars\thanks{Based on data obtained from the ESO Science Archive Facility.}}

\author{ F. Paletou\inst{1,2} \and M. Gebran\inst{3} \and E. R. Houdebine\inst{1,2,4}
  \and V. Watson\inst{1,2}
          }

          \institute{Universit\'e de Toulouse, UPS-Observatoire
            Midi-Pyr\'en\'ees, Irap, Toulouse, France
\and
          CNRS,  Institut de Recherche en Astrophysique et
            Plan\'etologie, 14 av. E. Belin, F--31400 Toulouse, France
            \email{fpaletou@irap.omp.eu}
\and
         Department of Physics and Astronomy, Notre Dame University-Louaize, PO Box 72, Zouk Mikaël, Lebanon
\and
         Armagh Observatory, College Hill, BT61 9DG Armagh, Northern Ireland
}

   \date{Received June 24, 2015; accepted July 14, 2015.}


   \abstract{ We show how the range of application of the principal
     component analysis-based inversion method of Paletou et
     al. (2015) can be extended to late-type stars data.  Besides
     being an extension of its original application domain, for FGK
     stars, we also used synthetic spectra for our learning database.
     We discuss our results on effective temperatures against previous
     evaluations made available from Vizier and Simbad services at
     CDS.}


   \titlerunning{Late-type stars effective temperatures}

   \maketitle

%

\section{Introduction}

Effective temperatures of late-type stars were inverted from HARPS
spectra, using the principal component analysis-based (PCA) method
detailed in Paletou et al. (2015). In the latter study, fundamental
parameters of FGK stars were inverted using a so-called learning
database made from the Elodie stellar spectra library (see Prugniel et
al. 2007) i.e., using \emph{observed} spectra for which fundamental
parameters were already evaluated. Also, spectra considered in Paletou
et al. (2015) study had typical spectral resolutions ${\cal R}$ of
50\,000 (Allende Prieto el al. 2004) and 65\,000 (Petit et al. 2014)
i.e., values significantly \emph{lower} than HARPS data.

In this study, the inversion of the effective temperature, $T_{\rm
  eff}$, from spectra of late-type (dwarf) stars of K and M spectral
types are made using a database of \emph{synthetic} spectra. We
discuss hereafter comparisons with published values, collected from
the VizieR and Simbad services of CDS.

\section{The learning database}

A grid of 6\,336 spectra was computed using {\sc Synspec}-48 synthetic
spectra code (Hubeny \& Lanz 1992) and Kurucz {\sc Atlas}-12 model
atmospheres (Kurucz 2005). The linelist was built from Kurucz (1992)
{\tt gfhyperall.dat}\footnote{http://kurucz.harvard.edu}.

For our purpose, we adopted indeed a grid of parameters such that
$T_{\rm eff}$ is in a 3500--4600 K range with a 100 K step, log$g$ is
in the range of 4--5 dex with a 0.2 dex step, metallicity [Fe/H] is in a
-2--+0.5 range with a 0.25 dex step and, finally, $v{\rm sin}i$ varies
from 0 to 14 $ {\rm km\,s}^{-1}$ with a 2 $ {\rm km\,s}^{-1}$
step.

For all models the microturbulent velocity was fixed at $\xi_t = 1
{\rm km\,s}^{-1}$ and [$\alpha$/Fe] was set to 0. The spectral
resolution of the HARPS spectrograph i.e., ${ {\cal R} } = 115\,000$
was adopted for the production of this set of synthetic spectra. We
finally limited the study to a spectral band centered around the
Na\,{\sc i} D-doublet, ranging from 585.3 to 593.2 nm.

Even though the choice of Atlas for such cool stars, as well as the
one of the spectral band we considered may be arguable, we show that
effective temperatures we inverted from HARPS data are realistic
enough for several further studies (e.g., rotation-activity
correlations vs. the spectral type).


\section{Observational and reference data}

We used 57 high-resolution HARPS spectra taken from the ESO Science
Archive Facility. The selection targeted late dwarf K stars, and early dwarf
M stars. High signal-to-noise spectra were also privileged. Related
objects are listed in Fig.\,(1) -- see also Table (1).

\begin{figure*}
  \includegraphics[width=19cm,angle=0]{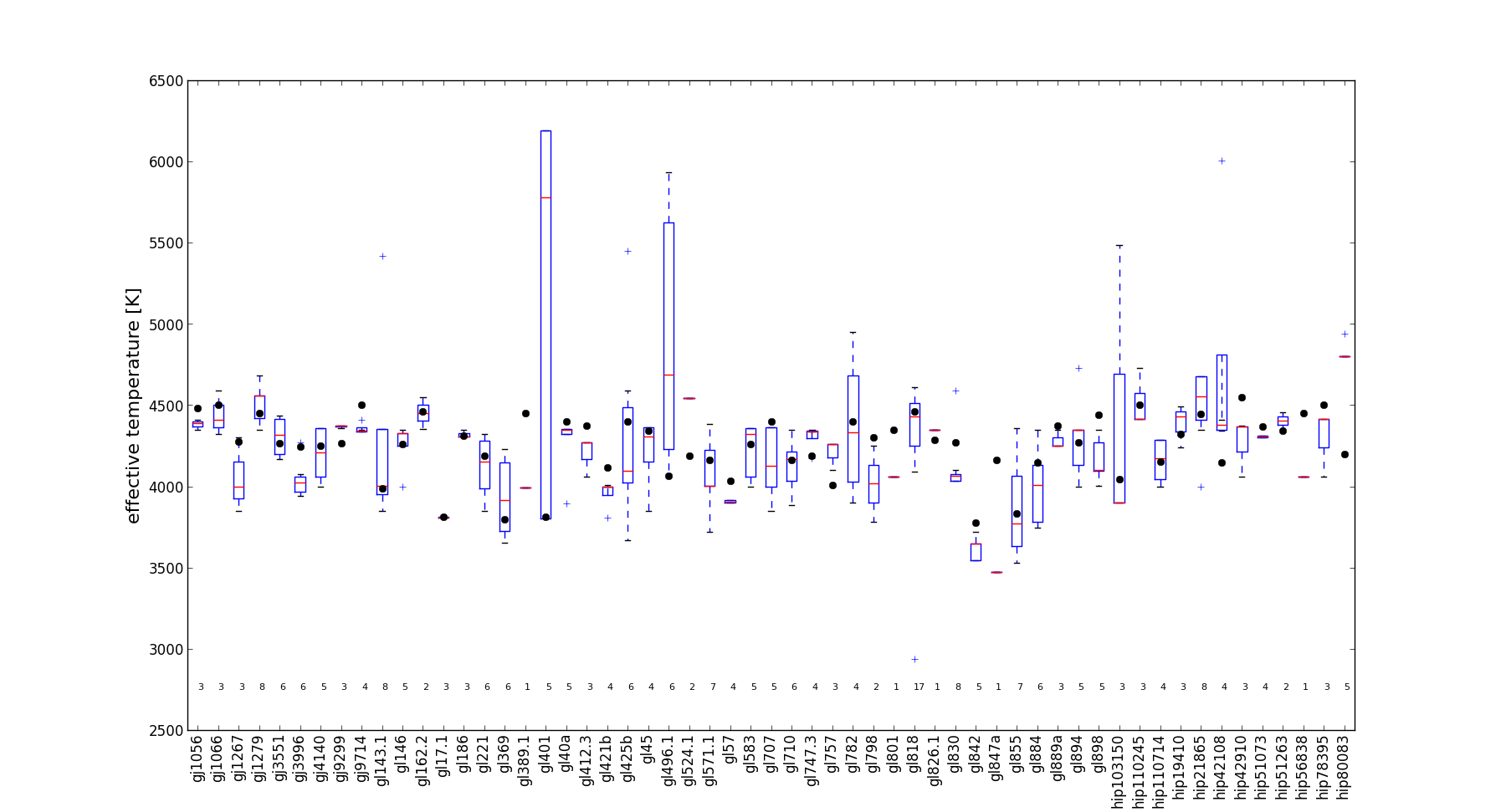}
  \caption{Comparison between our estimate of effective temperatures
    ($\bullet$), and the values we got from available VizieR
    catalogues. The latter collections are represented as classical
    boxplots (We remind here that the horizontal bar inside each
      box indicates the median, or $Q_2$ value, while each box extends
      from first quartile, $Q_1$, to third quartile $Q_3$. Extreme
      values but still within a 1.5 times the interquartile range away
      from either $Q_1$ or $Q_3$ are connected to the box with
      dashed-lines. Outliers are denoted by a '+' symbol). Objects we
    studied are listed along the horizontal axis. In addition, for
    each object at level $T_{\rm eff}\sim 2\,800$ K, we explicited the
    number of values found among all VizieR catalogues. }
  \label{Fig1}
\end{figure*}

All reference data were collected from VizieR catalogues, using the
{\sc Astroquery}\footnote{{\tt astroquery.readthedocs.org}} Python
modules already mentioned by Paletou \& Zolotukhin (2014).


\section{Results}

Given the ``bulk'' of nearest neighbours we consider for our inversion
procedure, we estimate that 150 K is a typical upper value for the
uncertainty on our derived effective temperature.

From Fig. (1), we can identify 8 major outliers considering deviations
from our estimate of the effective temperature and values published
and made available through the VizieR service at CDS.

GL\,401 is a M3V star. It is also an interesting case, from the point
of view of fundamental parameters made available. We could retrieve 5
determinations of its effective temperature, spanning an impressive
range of values, as can be seen from the corresponding boxplot in
Fig\,(1). However, our estimate of 3\,810 K is in excellent agreement
with the recent value of 3\,804 K derived by Gaidos et al. (2014).

The case of the K9V star GL\,496.1 is quite similar. We extracted 6
values from VizieR and again, the most recent value of 4\,075 K given
Gaidos et al. (2014) is in excellent agreement with our estimate of
4\,063 K. We also note the quite different value of 4\,685 K given by
Santos et al. (2013), and reported at Simbad@CDS for this object.

GL\,389.1 is believed to be a K5.5V star (Gray et al. 2006). We could
identify only one data at VizieR, an estimate of 3\,990 K given by
Lafrasse et al. (2010, also catalogue id. II/320). Our own estimate is
significantly hotter at 4\,450 K, which appears as more consistent with
the spectral type found at Simbad. 

According to Simbad@CDS, GL\,847\,A is a K4 star (Van Leeuwen
2007). This is clearly not consistent with our estimate of an
effective temperature of 4\,160 K. Neither with the only value we
could retrieve from VizieR i.e., the estimate of Morales et al. (2008)
giving 3\,470 K. As for the precedent object, the lack of data available
in catalogs makes difficult the assessment of a reliable reference value.

HIP\,42108 is a K6V star (Gray et al. 2006) for which we could
estimate an effective temperature of 4\,147 K. From available
catalogs, we could find the most recent estimate of 4\,343 K given by
McDonald et al. (2012), also very close to the alternative estimate of
Wright et al. (2003). Lafrasse et al. (2010) report also a quite close
estimate of 4\,410 K. However for this object, Ammons et al. (2006)
report a very different value of 6\,005 K.

A similar case of overestimation has been identified for GL\,524.1. We
obtained indeed an effective temperature of 4\,186 K, while Ammons et
al. (2006) provide a significantly larger value of 4\,554 K. We could
not find alternative estimates for this parameter and object
unfortunately (there is an entry on GL\,524.1 in the so-called
\emph{Distances and atmospheric parameters of MSU stars} of Morales et
al. 2008, but no $T_{\rm eff} $ value is given).

HIP\,56838 is, according to Simbad, a K6V star after the contribution
of Gray et al. (2006). Our estimate of its effective temperature is
4\,200 K, in agreement with a K6 (main sequence) spectral type, while
the only VizieR data we could get is 4\,060 K (Wright et
al. 2003). Even though this object can also be found in Ammons et
al. (2006) catalogue, no $T_{\rm eff}$ value is given there.
Besides, a much cooler temperature of 3\,800 K that would better
correspond to a M0V spectral type was recently given by Kordopatis et
al. (2013).

HIP\,80083 is quite consistently given at 4\,800 K (Sousa et al. 2011,
Adibekyan et al. 2012, Carretta 2013), or even hotter (Ammons et
al. 2006). Our inversion procedure gives an effective temperature
significantly lower at 4\,200 K, typical of a spectral type later than
K4, as indicated by Simbad@CDS.

Table (1) summarizes our results. It displays our inverted $T_{\rm
     eff}^{(\rm inv.)}$, and two reference values.  The first one,
   $T_{\rm eff}^{(\rm clos.)}$, was defined as the value found in
   VizieR catalogues closest to our estimate, while $T_{\rm eff}^{(\rm
     med.)}$ is the \emph{median} of all catalogue values.

   Finally, in order to characterize our results, we first removed
   from our objects list the 8 above\-mentioned outliers, that is
   about 14\% of the original sample. Considering reference values as
   the one closest to our inverted effective temperature, we obtain a
   (mean signed difference or) bias of 21 K and a standard deviation
   of 90 K. Should we use the median value as reference, bias is 60 K
   and standard deviation 132 K.


\begin{table}
  \caption{Inverted and reference effective temperatures for all objects. 
    Hereafter $T_{\rm eff}^{(\rm clos.)}$ was defined as
    the value found in VizieR catalogues closest to our inverted
    $T_{\rm eff}^{(\rm inv.)}$, while $T_{\rm eff}^{(\rm med.)}$ is the
    median of catalogues values.}
\label{table:1}
\centering
\begin{tabular}{cccc}
  \hline\hline
  Object & $T_{\rm eff}^{(\rm inv.)}$[K] & $T_{\rm eff}^{(\rm
    clos.)}$[K] & $T_{\rm eff}^{(\rm med.)}$[K]
  \\
  \hline
   gj1056  &   4480.0  &   4410.0  &   4391.0 \\
   gj1066  &   4500.0  &   4590.0  &   4410.0 \\
   gj1267  &   4275.0  &   4301.0  &   4000.0 \\
   gj1279  &   4450.0  &   4424.0  &   4556.0 \\
   gj3551  &   4267.0  &   4287.0  &   4318.5 \\
   gj3996  &   4244.0  &   4272.0  &   4023.0 \\
   gj4140  &   4250.0  &   4210.0  &   4210.0 \\
   gj9299  &   4267.0  &   4360.0  &   4372.0 \\
   gj9714  &   4500.0  &   4410.0  &   4344.5 \\
  gl143.1  &   3985.0  &   3970.0  &   4001.0 \\
    gl146  &   4260.0  &   4250.0  &   4329.0 \\
  gl162.2  &   4462.5  &   4550.0  &   4452.5 \\
   gl17.1  &   3812.5  &   3809.0  &   3809.0 \\
    gl186  &   4311.0  &   4307.0  &   4307.0 \\
    gl221  &   4186.0  &   4282.0  &   4151.0 \\
    gl369  &   3796.0  &   3915.0  &   3915.0 \\
  gl389.1  &   4450.0  &   3990.0  &   3990.0 \\
    gl401  &   3810.0  &   3804.0  &   5780.0 \\
    gl40a  &   4400.0  &   4352.0  &   4350.0 \\
  gl412.3  &   4375.0  &   4271.0  &   4271.0 \\
   gl421b  &   4117.0  &   4008.0  &   3995.0 \\
   gl425b  &   4400.0  &   4590.0  &   4097.5 \\
     gl45  &   4340.0  &   4361.0  &   4305.5 \\
  gl496.1  &   4063.0  &   4075.0  &   4685.0 \\
  gl524.1  &   4186.0  &   4544.0  &   4544.0 \\
  gl571.1  &   4164.0  &   4060.0  &   4002.0 \\
     gl57  &   4033.0  &   3913.0  &   3906.5 \\
    gl583  &   4261.0  &   4320.0  &   4320.0 \\
    gl707  &   4400.0  &   4364.0  &   4125.0 \\
    gl710  &   4160.0  &   4130.0  &   4165.0 \\
  gl747.3  &   4186.0  &   4170.0  &   4337.0 \\
    gl757  &   4008.0  &   4100.0  &   4259.0 \\
    gl782  &   4400.0  &   4590.0  &   4330.0 \\
    gl798  &   4300.0  &   4250.0  &   4015.5 \\
    gl801  &   4350.0  &   4060.0  &   4060.0 \\
    gl818  &   4462.5  &   4444.0  &   4430.0 \\
  gl826.1  &   4287.5  &   4350.0  &   4350.0 \\
    gl830  &   4269.0  &   4100.0  &   4065.5 \\
    gl842  &   3776.0  &   3720.0  &   3649.0 \\
   gl847a  &   4160.0  &   3470.0  &   3470.0 \\
    gl855  &   3831.0  &   3771.0  &   3771.0 \\
    gl884  &   4147.0  &   4130.0  &   4009.5 \\
   gl889a  &   4371.0  &   4350.0  &   4251.0 \\
    gl894  &   4270.0  &   4350.0  &   4350.0 \\
    gl898  &   4440.0  &   4350.0  &   4101.0 \\
hip103150  &   4043.0  &   3900.0  &   3900.0 \\
hip110245  &   4500.0  &   4416.0  &   4416.0 \\
hip110714  &   4150.0  &   4060.0  &   4172.0 \\
 hip19410  &   4320.0  &   4238.0  &   4432.0 \\
 hip21865  &   4445.5  &   4432.0  &   4555.0 \\
 hip42108  &   4147.0  &   4343.0  &   4380.0 \\
 hip42910  &   4550.0  &   4373.0  &   4368.0 \\
 hip51073  &   4367.0  &   4310.0  &   4305.0 \\
 hip51263  &   4343.0  &   4350.0  &   4403.5 \\
 hip56838  &   4450.0  &   4060.0  &   4060.0 \\
 hip78395  &   4500.0  &   4414.0  &   4414.0 \\
 hip80083  &   4200.0  &   4800.0  &   4800.0 \\
  \hline
\end{tabular}
\end{table}

\section{Conclusion}

We have shown that the PCA-based inversion method of Paletou et
al. (2015) provides \emph{realistic} values for the effective
temperature of late-type stars. Comparisons made between our estimates
and effective temperature data found in the available literature
reveals the existence of some strong discrepancies for a few
objects. The latter are most often related to very limited samples of
estimates, so that additional investigations are clearly required for
these objects.

These should consist in using different synthetic spectra that can
produce other radiative modelling tools such as Marcs (Gustafsson et
al. 2008) or Phoenix, for cool stars (see e.g., Husser et
al. 2013). The consideration of \emph{other} spectral domains, and
eventually the use of a \emph{combination} of several distinct
spectral domains should be explored too.

Our study rises also the more general question of the consistency
between published (and made available) data, as well as the
consistency between data respectively provided by VizieR and Simbad
services.

\begin{acknowledgements}
  This research has made use of the VizieR catalogue access tool, CDS,
  Strasbourg, France. The original description of the VizieR service
  was published in A\&AS 143, 23. This research has made use of the
  SIMBAD database, operated at CDS, Strasbourg, France. \emph{Per
    Ardua, ad Astra\,!}
\end{acknowledgements}


\begin{thebibliography}{}

\bibitem[2012]{adibekyan} Adibekyan, V. Zh., Sousa, S. G., Santos,
  N. C., Delgado Mena, E., González Hernández, J. I., Israelian, G.,
  Mayor, M., Khachatryan, G., 2012, \aap, 545, A32

\bibitem[2004]{s4n} Allende Prieto, C., Barklem, P.S., Lambert, D.L.,
  Cunha, K. 2004, \aap, 420, 183

\bibitem[2006]{ammons} Ammons, S. M., Robinson, S. E., Strader, J.,
  Laughlin, G., Fischer, D., Wolf, A., 2006, \apj, 638, 1004

\bibitem[2013]{carretta} Carretta, E., 2013, \aap, 557, 128

\bibitem[2014]{gaidos} Gaidos, E., Mann, A. W., Lépine, S., Buccino,
  A., James, D., Ansdell, M., Petrucci, R., Mauas, P., Hilton,
  E. J., 2014, \mnras, 443, 2561

\bibitem[206]{gray} Gray, R. O., Corbally, C. J., Garrison, R. F.,
  McFadden, M. T., Bubar, E. J., McGahee, C. E., O'Donoghue, A. A.,
  Knox, E. R., 2006, \aj, 132, 161

\bibitem[2008]{marcs} Gustafsson B., Edvardsson B., Eriksson K.,
  Jørgensen U.G., Nordlund Å., Plez B., 2008, \aap, 486, 951.

\bibitem[1992]{synspec} Hubeny, I., Lanz, T. 1992, \aap, 262, 501

\bibitem[2013]{phoenix} Husser, T.-O., Wende-von Berg, S., Dreizler,
  S., Homeier, D., Reiners, A., Barman, T., Hauschildt, P. H., 2013,
  \aap, 553, A6

\bibitem[2013]{kordopatis}  Kordopatis et al., 2013, \aj, 146, 134

\bibitem[1992]{kurucz1992} Kurucz, R. L. 1992, Rev. Mex. Astron. Astrofis., 23, 45

\bibitem[2005]{kurucz2005} Kurucz, R. L. 2005, Mem. Soc. Astron. Ital. Supp., 8, 14

\bibitem[2010]{lafrasse}  Lafrasse S., Mella G., Bonneau D., Duvert G., Delfosse X., Chelli A., 2010,
SPIE Conf. on Astronomical Instrumentation, 77344E, 140

\bibitem[2012]{mcdo} McDonald, I., Zijlstra, A. A., Boyer, M. L.,
  2012, \mnras, 427, 343

\bibitem[2008]{morales} Morales, J. C., Ribas, I., Jordi, C., 2008, \aap, 478, 507

\bibitem[2014]{paletouzolo2014} Paletou, F., Zolotukhin, I., 2014 [{\tt
    arXiv:1408.7026}] 
	 
\bibitem[2015]{paletou2015} Paletou, F., B\"ohm, T., Watson, V.,
  Trouilhet, J.-F., 2015, \aap, 573, A67

\bibitem[2014]{polarbase} Petit, P., Louge, T., Th\'eado, S., et
  al. 2014, \pasp, 126, 469


\bibitem[2007]{prugniel} Prugniel, P., Soubiran, C., Koleva, M., Le
  Borgne, D., 2007, {\tt[arXiv:astro-ph/0703658]}

\bibitem[2013]{santos} Santos, N. C., Sousa, S. G., Mortier, A.,
  Neves, V., Adibekyan, V., Tsantaki, M., Delgado Mena, E., Bonfils,
  X., Israelian, G., Mayor, M., Udry, S., 2013, \aap, 556, A150

\bibitem[2011]{sousa} Sousa, S. G., Santos, N. C., Israelian, G.,
  Mayor, M., Udry, S., 2011, \aap, 533, A141

\bibitem[2007]{leeuwen}  van Leeuwen, F., 2007, \aap, 474, 653

\bibitem[2003]{wright} Wright, C. O., Egan, M. P., Kraemer,
K. E., Price, S. D., 2003, \aj, 125, 359
 
\end{thebibliography}
\end{document}